\documentclass[a4paper]{jpconf}
\usepackage{graphicx}
\begin{document}
\title{A Photometric Analysis of ZZ Ceti Stars: A Parameter-Free Temperature
Indicator?}

\author{P Bergeron$^1$, S K Leggett$^2$ and H C Harris$^3$}

\address{$^1$ D\'epartement de Physique, Universit\'e de Montr\'eal, C.P.~6128, Succ.~Centre-Ville, Montr\'eal, Qu\'ebec H3C 3J7, Canada}
\address{$^2$ Gemini Observatory, Northern Operations Center, 670 North A'ohoku Place, Hilo, Hawaii 96720, USA}
\address{$^3$ US Naval Observatory, Flagstaff Station, Flagstaff, Arizona 86001, USA}

\ead{bergeron@astro.umontreal.ca, sleggett@gemini.edu, hch@nofs.navy.mil}

\begin{abstract}
We present a model atmosphere analysis of optical $VRI$ and infrared $JHK$
photometric data of about two dozen ZZ Ceti stars. We first show from
a theoretical point of view that the resulting energy distributions
are not particularly sensitive to surface gravity or to the assumed
convective efficiency, a result which suggests a parameter-free
effective temperature indicator for ZZ Ceti stars. We then fit the
observed energy distributions with our grid of model atmospheres and
compare the photometric effective temperatures with the spectroscopic
values obtained from fits to the hydrogen line profiles. Our results
are finally discussed in the context of the determination of the
empirical boundaries of the ZZ Ceti instability strip.
\end{abstract}

\section{Introduction}

The atmospheres of hydrogen-line (DA) white dwarfs become convective
below roughly $T_{\rm eff}\sim17,000$~K, although it is only when they
have cooled below $\sim 12,000$~K that a significant fraction of
the total flux is transported by convection. At that point, the bottom
of the convection zone sinks deeper into the star until it reaches a
maximum depth of $\Delta M/M_\star\sim 10^{-6}$ near $T_{\rm eff}\sim
5000$~K. Below 9000~K or so, convection in hydrogen-atmosphere white
dwarfs becomes adiabatic, and the thermodynamic stratification is then
specified by the value of the adiabatic gradient and the hydrostatic
equilibrium equation. As such, the convective flux becomes almost
completely independent of the assumed convective efficiency. Above
that temperature, however, the structure of the atmosphere and the
deeper stellar envelope depend sensitively on the way convection is
treated in the model calculations.  For instance, Bergeron et
al.~(1992) have shown that in the range $T_{\rm eff}=10,000-14,000$~K,
the predicted energy distributions and line profiles are strongly
affected by the parameterization of the mixing-length theory,
the most commonly used formalism that can take into account convective
energy transport in model atmosphere calculations.

This is unfortunately the precise range of effective temperature where
the DA white dwarf pulsators, or ZZ Ceti stars, are found. According
to the spectroscopic analysis of Gianninas et al.~(2006, see also
these proceedings), ZZ Ceti stars occupy a trapezoidal region in the
$T_{\rm eff}-\log g$ plane, with an empirical blue edge around
12,400~K and a red edge near 11,000~K, although both edges show a
strong dependence on surface gravity. It must be stressed, however,
that the atmospheric parameter determinations of ZZ Ceti stars depend
strongly on the assumed convective efficiency in the model atmosphere
calculations. Moreover, this efficiency cannot be easily estimated
from first principles. To overcome this situation, Bergeron et
al.~(1995) have attempted to calibrate the convective efficiency in
the atmospheres of DA stars by comparing effective temperatures
obtained from fits to Balmer line profiles with those derived from UV
energy distributions. The authors convincingly demonstrate that the
so-called ML2/$\alpha=0.6$ parametrization of the mixing-length theory
provides the best internal consistency between optical and UV
effective temperatures, as well as trigonometric parallaxes, $V$
magnitudes, and gravitational redshifts. This has been the
parametrization used in all model atmosphere calculations since then.

In this paper, we present an independent method for estimating the
internal consistency of this parametrization by fitting optical $VRI$
and infrared $JHK$ photometric energy distributions of ZZ Ceti stars
with detailed model atmosphere calculations. In particular, we show
that this technique provides a temperature indicator for ZZ Ceti stars
that is independent of the assumed convective efficiency.

\section{A Parameter-Free Temperature Indicator}

Our fitting technique relies on the approach originally developed by
Bergeron et al.~(1997) in their study of cool white dwarfs. Briefly,
the atmospheric parameters for each star are obtained by converting
the optical and infrared photometry into observed fluxes, and by
comparing the resulting spectral energy distributions with those
predicted from model atmosphere calculations. The first step is
accomplished by transforming the magnitudes into average stellar
fluxes $f_{\lambda}^m$ received at Earth using the calibration of
Holberg et al.~(2006). The observed and model fluxes, which depend on
$T_{\rm eff}$ and $\log g$, are related by the equation
$f_{\lambda}^m= 4\pi~(R/D)^2~H_{\lambda}^m$ where $R/D$ is the ratio
of the radius of the star to its distance from Earth, and
$H_{\lambda}^m$ is the Eddington flux, properly averaged over the
corresponding filter bandpass. We finally minimize the difference
between observed and predicted fluxes at all bandpasses using our
standard Levenberg-Marquardt minimization procedure. Only $T_{\rm
eff}$ and the solid angle $\pi~(R/D)^2$ are considered free
parameters.

As a theoretical exploration of our method, we show in Figure 1 the predictions from our model atmospheres
for various values of $T_{\rm eff}$, $\log g$, and convective
efficiency. Monochromatic Eddington fluxes are shown together with the
average over the optical $UBVRI$ and infrared $JHK$ filter bandpasses.
One can already notice that if the $U$ filter, and to a lesser extent,
the $B$ filter are omitted from the fits, the predicted energy distributions
(normalized at $V$) do not depend on the assumed value of $\log g$
(middle panel), and more interestingly, they do not depend on the assumed
parametrization of the mixing-length theory either (top panel; the
ML1, ML2, and ML3 prescriptions are described in detail in Bergeron et
al.~1995). Fortunately, the energy distributions remain sensitive to
variations of effective temperature (bottom panel). Hence, by restricting our
analysis to optical $VRI$ and infrared $JHK$ photometry, we obtain an
{\it independent temperature scale} for ZZ Ceti stars that does not
depend on the assumed value of $\log g$ (or mass) or convective efficiency.

\begin{figure}[h]
\centering
\includegraphics[width=26pc]{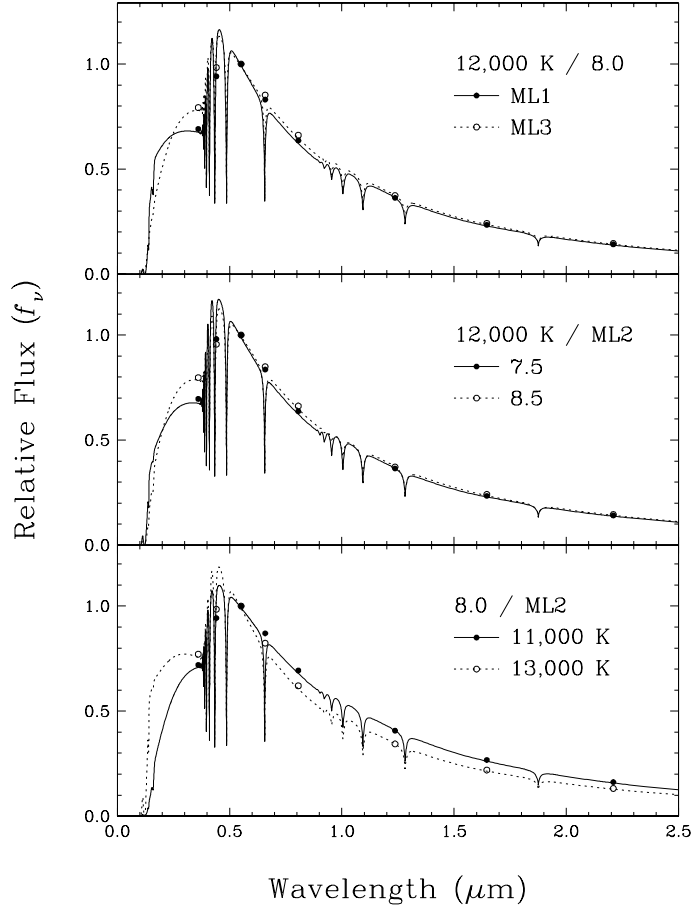}
\caption{Energy distributions for white dwarfs in the ZZ Ceti temperature range
for various values of $T_{\rm eff}$, $\log g$, and convective
efficiency. Also shown (open and filled circles) are the integrated
fluxes over the optical $UBVRI$ and infrared $JHK$ filter bandpasses. All
distributions are normalized at $V$.}
\end{figure}

\begin{figure}[h]
\centering
\includegraphics[width=34pc]{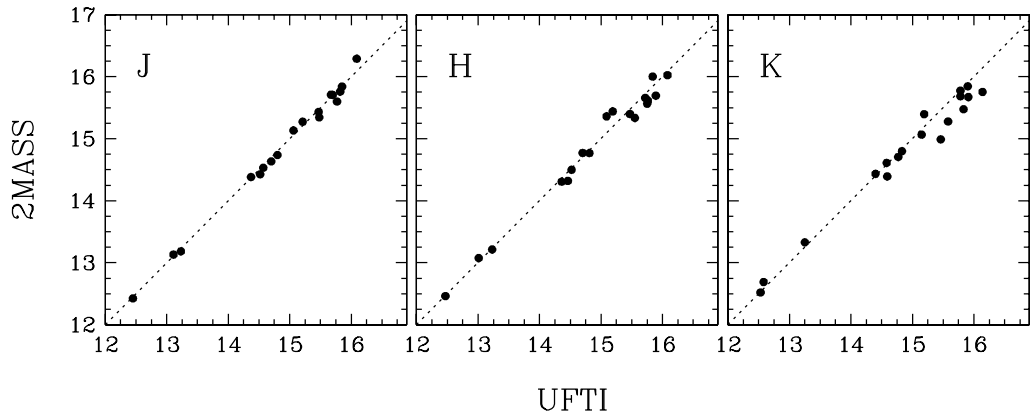}
\caption{Comparison of infrared photometry for 17 ZZ Ceti stars in our sample 
that have both UFTI and 2MASS $JHK$ measurements.}
\end{figure}

\section{Observational data}

Since 2001, we have been obtaining time-averaged optical and infrared
photometric observations of ZZ Ceti stars. Our data for 24 ZZ Ceti
white dwarfs are summarized in Table 1.  The optical photometry were
secured with a CCD detector attached to the 1.0 m R/C reflector at the
U.S.~Naval Observatory in Flagstaff. Data were calibrated by observing
standards from Landolt (1992) throughout each night and determining
nightly extinction and color terms.  When possible, stars were
observed on two to three different nights to average out their
variability. We measured the $V$ magnitude for all 24 objects in our
sample, and $R$ and $I$ magnitudes for only 9 and 20 objects,
respectively. The infrared $JHK$ photometry was secured for 19 objects
using the U.K. Infrared Telescope (UKIRT) Fast-Track Imager (UFTI,
Roche et al.~2003). The camera's filters are on the Mauna Kea
Observatories photometric system (Tokunaga et al.~2002), and the data
were calibrated using standards from Leggett et al.~(2006). Since the
beginning of our survey, 2MASS photometry has also become available
for 23 objects in our sample, and this data is reported in Table 1 and
included in our analysis below. The UFTI and 2MASS $JHK$ photometric
observations for the 17 objects in common are compared in Figure
2. Both data sets agree extremely well, with the exception perhaps of
the $K$ filter for faint stars where our UFTI photometry is arguably
better.

\begin{table}
\caption{\label{table1}Optical and infrared photometry of ZZ Ceti stars.}
\begin{center}
\begin{tabular}{lccccccccc}
\br
&&& & & UFTI & & & 2MASS & \\ Name & V & R & I & J & H & K & J & H & K
\\
\mr
Ross 548        &   14.16 &       &       & 14.37 & 14.36 & 14.40 & 14.38 & 14.30 & 14.43\\
KUV 02464+3239  &   16.03 &       & 16.03 &       &       &       & 15.97 & 15.92 & 15.44\\
HL 76           &   14.99 &       & 15.05 & 15.06 & 15.09 & 15.15 & 15.13 & 15.36 & 15.06\\
G38-29          &   15.59 &       &       & 15.70 & 15.75 & 15.78 & 15.70 & 15.56 & 15.77\\
HS 0507+0435B   &   15.33 &       & 15.39 & 15.47 & 15.55 & 15.58 & 15.43 & 15.33 & 15.27\\
GD 66           &   15.55 &       & 15.61 & 15.67 & 15.72 & 15.78 & 15.70 & 15.65 & 15.68\\
KUV 08368+4026  &   15.60 & 15.62 & 15.64 & 15.85 & 15.89 & 15.90 & 15.84 & 15.69 & 15.84\\
GD 99           &   14.51 & 14.53 & 14.52 & 14.70 & 14.70 & 14.77 & 14.63 & 14.77 & 14.70\\
G117-B15A       &   15.46 & 15.51 & 15.51 & 15.77 & 15.76 & 15.83 & 15.59 & 15.61 & 15.47\\
KUV 11370+4222  &   16.55 & 16.57 & 16.56 & 16.78 & 16.76 & 16.68 &       &       &      \\
G255-2          &   15.97 & 15.95 & 16.00 &       &       &       & 16.00 & 15.87 & 16.28\\
GD 154          &   15.26 & 15.24 & 15.29 & 15.48 & 15.47 & 15.46 & 15.34 & 15.39 & 14.98\\
G238-53         &   15.46 & 15.52 & 15.52 &       &       &       & 15.68 & 15.41 & 15.30\\
EC 14012-1446   &   15.66 & 15.66 & 15.69 & 15.82 & 15.84 & 15.91 & 15.75 & 16.00 & 15.67\\
GD 165          &   14.27 & 14.33 & 14.34 & 14.57 & 14.52 & 14.58 & 14.53 & 14.49 & 14.61\\
PG 1541+651     &   15.54 &       & 15.58 &       &       &       & 15.60 & 15.91 & 15.42\\
G226-29         &   12.24 &       &       & 12.45 & 12.47 & 12.53 & 12.42 & 12.46 & 12.52\\
G207-9          &   14.61 &       & 14.63 & 14.80 & 14.81 & 14.83 & 14.73 & 14.76 & 14.79\\
G185-32         &   12.98 &       &       & 13.23 & 13.23 & 13.25 & 13.18 & 13.21 & 13.32\\
GD 385          &   15.10 &       & 15.12 & 15.21 & 15.19 & 15.19 & 15.27 & 15.44 & 15.39\\
G232-38         &   16.83 &       & 16.79 &       &       &       & 16.34 & 15.97 & 16.01\\
PG 2303+243     &   15.26 &       & 15.34 &       &       &       & 15.49 & 15.41 & 15.70\\
G29-38          &   13.04 &       & 13.01 & 13.11 & 13.01 & 12.58 & 13.13 & 13.07 & 12.68\\
G30-20          &   16.07 &       & 16.04 & 16.09 & 16.08 & 16.14 & 16.29 & 16.02 & 15.75\\
\br
\end{tabular}
\end{center}
\end{table}

\section{Photometric analysis}

As discussed above, the magnitudes from Table 1 are converted into
average fluxes using the appropriate filter transmission functions for
each photometric system. Similarly, the model monochromatic fluxes are
also averaged over the same filter sets. The model atmospheres used in
this analysis are slightly different from those used by Bergeron et
al.~(1995) and in all previous analyses of ZZ Ceti stars by the
Montreal group (Gianninas et al.~2006 and references therein).  We are
now making use of the extensive Stark broadening profiles calculated
by Lemke (1997), while our earlier calculations relied on similar
calculations by Sch\"oning \& Butler (1991, private communication) but
for temperatures and densities characteristic of hot subdwarfs stars,
which often required extrapolations in the higher density regime of
white dwarf atmospheres. With these new models in hand, we redid the
analysis of Bergeron et al.~(1995) and found that a slightly more
efficient version of the mixing-length theory, namely ML2 with
$\alpha=0.7$ (instead of 0.6), provides the best internal consistency
between effective temperatures obtained from optical and UV
spectra. Sample fits to the $VRI$ and $JHK$ energy distributions of
four ZZ Ceti stars in our sample using these models are displayed in
Figure 3. We assume in this figure a value of $\log g=8.0$ for all
objects.

\begin{figure}[t]
\centering
\includegraphics[width=28pc,angle=270]{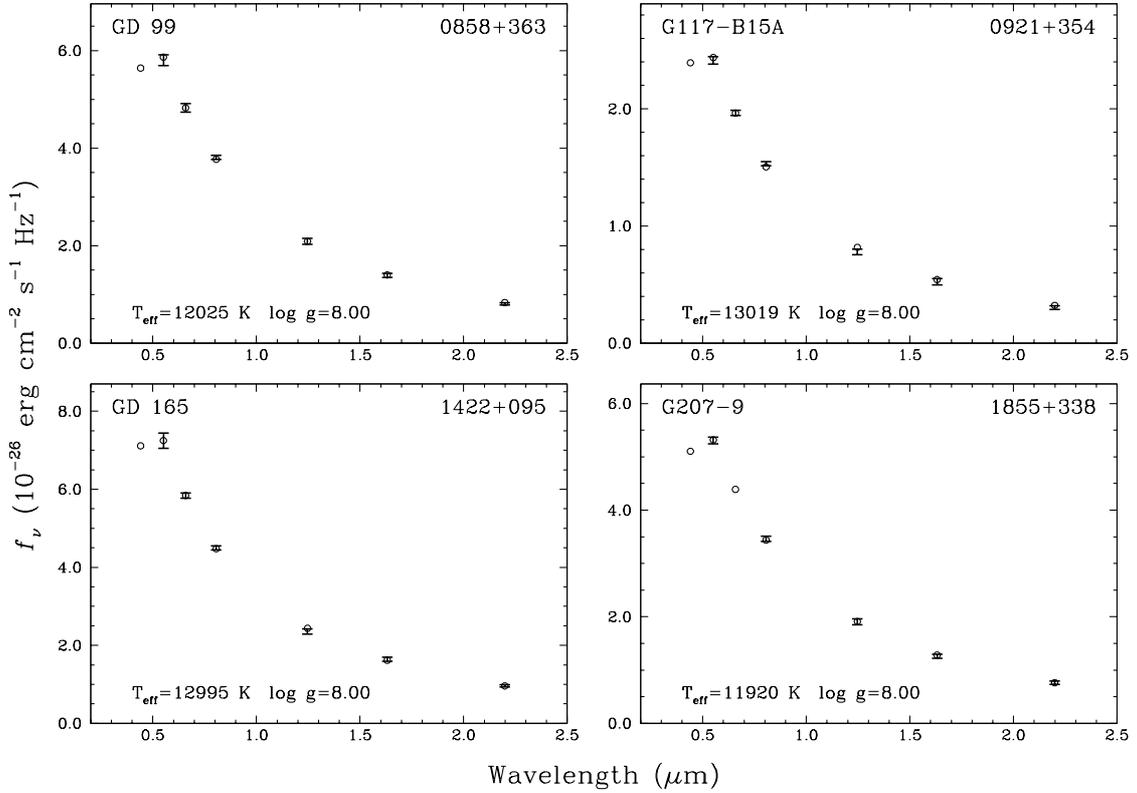}
\caption{Sample fits to the energy distributions of ZZ Ceti stars.
The $VRI$ and $JHK$ photometric observations are represented by error
bars while the model fluxes are shown as open circles.}
\end{figure}

The photometric effective temperatures for all ZZ Ceti stars in our sample are compared in
Figure 4 with spectroscopic temperatures obtained from fits to the
hydrogen Balmer lines using the so-called spectroscopic technique (see
Bergeron et al.~1995 and references therein); the optical spectra 
are taken from the analysis of
Gianninas et al.~(2006). Both infrared data sets (UFTI and 2MASS) are
considered independently in Figure 4. We also assume for our
photometric fits a value of $\log g=8.1$ for all objects, which
corresponds to the average spectroscopic value of our sample
($\langle\log g\rangle=8.143$, $\sigma=0.166$). A variation of $\log
g$ by $\pm 0.25$ dex from the mean value changes the photometric
temperatures by only 200~K, on average, as expected from the results
displayed in the middle panel of Figure 1. Overall, the agreement
observed in Figure 4 is satisfactory, although there are some objects
with photometric temperatures significantly larger than the
spectroscopic estimates. For these stars, however, we also notice
that the photometric uncertainties appear larger. The object all the way to the left in the 2MASS panel is G29-38, a ZZ
Ceti star with an infrared excess due to the presence of a dust disk,
which affects our photometric temperature determination (the corresponding measurement with
the UFTI photometry is outside the temperature range displayed here).

\begin{figure}[t]
\centering
\includegraphics[width=19pc,angle=270]{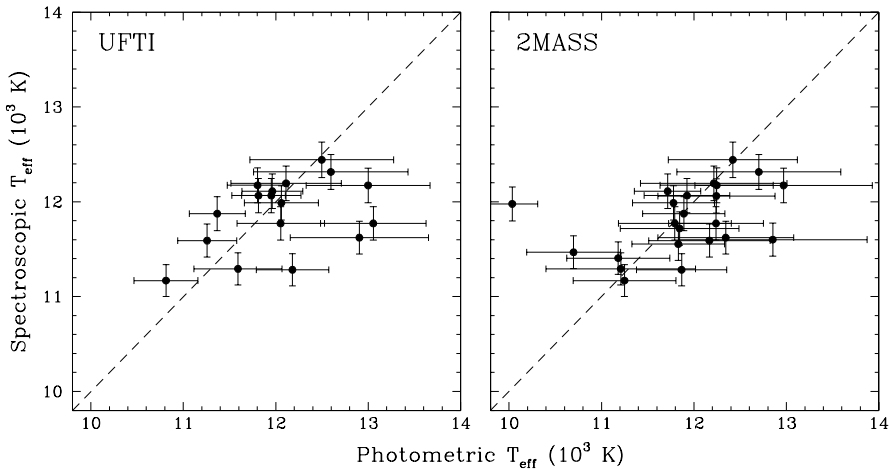}
\caption{Comparison of photometric and spectroscopic effective temperatures for
the two sets of infrared photometry used in our analysis. Model
atmospheres calculated with the ML2/$\alpha=0.7$ version of the
mixing-length theory are used to obtain both estimates of $T_{\rm
eff}$.}
\end{figure}

The photometric temperatures in Figure 4 are spread over a somewhat wider range 
of values than the spectroscopic temperature determinations. This can be 
explained by the fact that the photometric method is less accurate
than the spectroscopic method in this particular range of effective temperatures.
For instance, a difference of $\Delta T_{\rm eff}=2000$~K yields energy
distributions (bottom of Fig.~1) that differ less than the predicted
strengths of the Balmer line profiles for the same difference in $T_{\rm eff}$.

Of greater interest in the context of our study is the comparison of
photometric and spectroscopic effective temperatures for different
versions of the mixing-length theory. This is shown in Figure 5 for
the ML1, ML2, and ML3 parameterizations (in what follows we consider the
UFTI and 2MASS infrared data as independent measurements and simply merge the samples). Although the
spectroscopic $T_{\rm eff}$ values vary over a range of nearly 4000~K
going from ML1 (less efficient version) to ML3 (more efficient version), the
photometric temperatures remain nearly unchanged, as anticipated from our
results shown in the top panel of Figure 1. Also, for both temperature
estimates to agree, one must rely on a parametrization which is less
efficient than the ML2/$\alpha=1.0$ of the mixing-length
theory (used to produce Fig.~5), in agreement with the lower value 
of $\alpha=0.7$ adopted in the
comparison shown in Figure 4, which is reproduced in Figure 6 by
combining both UFTI and 2MASS data sets. These results add additional
confidence to our conclusion that the atmospheric convective
efficiency in DA white dwarfs is fairly well constrained and
understood.

\begin{figure}[h]
\centering
\includegraphics[width=25pc]{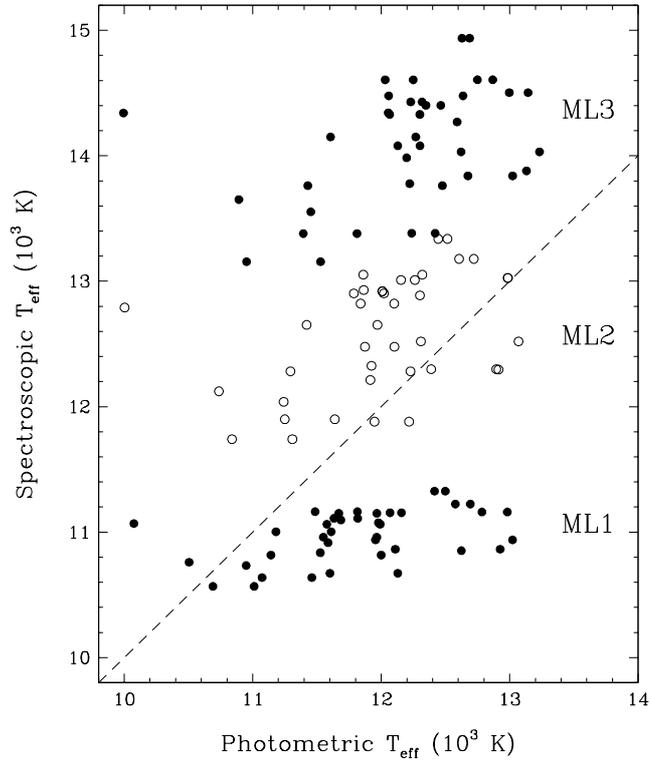}
\caption{Comparison of photometric and spectroscopic effective temperatures for
different versions of the mixing-length theory.}
\end{figure}

\begin{figure}[h]
\centering
\includegraphics[width=21pc]{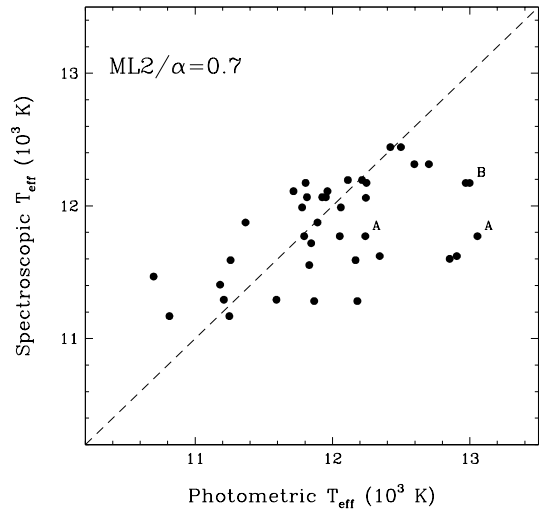}
\caption{Comparison of photometric and spectroscopic effective temperatures obtained
with model atmospheres calculated with ML2/$\alpha=0.7$. A value of
$\log g=8.1$ was used for the photometric determinations. The objects
labeled in the figure and discussed in the text correspond to (A) G117-B15A and
(B) GD 165.}
\end{figure}

Despite these encouraging results, there are still some stars in
Figure 6 for which the differences in photometric and spectroscopic
temperatures are uncomfortably large. One example labelled A in the
figure is G117-B15A. The spectroscopic solution yields $T_{\rm
eff}=11,770$~K and $\log g=8.04$ while the photometric solution with
the UFTI infrared data is about 1250~K hotter. On the other hand, the
photometric solution obtained with the 2MASS data is only 400~K
hotter. We believe this reflects the difficulty with obtaining precise
$T_{\rm eff}$ measurements for individual objects from photometry in
this particular range of effective temperature due mostly to the
sensitivity of the method, but also to problems with the calibration
of the synthetic photometry. An alternative, perhaps more exotic
explanation is provided by the second object, GD 165, labelled B in 
Figure 6. For this star, the spectroscopic solution yields $T_{\rm
eff}=12,170$~K and $\log g=8.13$. In this case, however, both
solutions with the UFTI and 2MASS photometry agree within 90~K, with
an average value of $T_{\rm eff}=12,950$~K, which is nearly 800~K
hotter than the spectroscopic solution. Interestingly enough, had we
used the ML2/$\alpha=1.0$ version of the mixing-length theory, both
spectroscopic and photometric temperatures would have been in perfect
agreement (the spectroscopic temperature increases to a value of
$T_{\rm eff}=13,030$~K in this case). Since GD 165 is one of the hottest ZZ Ceti star,
perhaps our results is suggesting that convective energy transport is
more efficient near the blue edge of the ZZ Ceti instability strip.

\section{Conclusion}

We have successfully demonstrated  that energy
distributions of ZZ Ceti stars built from optical $VRI$ and infrared
$JHK$ photometric data provide a parameter-free and independent
temperature scale for these stars that does not depend on the assumed
value of $\log g$ or convective efficiency. Although we believe that
the effective temperatures measured from photometry are less accurate
for {\it individual} objects than temperatures obtained from the
standard spectroscopic technique, our results nevertheless show that
a parameterization of the mixing-length theory with ML2/$\alpha=0.7$
provides, on average, an excellent internal consistency between
photometric and spectroscopic temperatures. In particular, both
methods yield consistent values for the boundaries of the ZZ Ceti
instability strip between roughly $T_{\rm eff}=11,000$~K and 12,500~K.

\ack{This work was supported in part by the NSERC Canada and by the Fund FQRNT
(Qu\'ebec). P. Bergeron is a Cottrell Scholar of Research Corporation
for Science Advancement. The United Kingdom Infrared Telescope is operated by the Joint Astronomy Centre on behalf of the Science and Technology Facilities Council of the U.K.}

\section*{References}

\begin{thereferences}

\item Bergeron P, Ruiz M  T, and Leggett S K 1997 {\sl ApJ Suppl} {\bf 108} 339
\item Bergeron P, Wesemael F, and Fontaine G 1992 {\sl ApJ} {\bf 387}, 288
\item Bergeron P, Wesemael F, Lamontagne R, Fontaine G, Saffer R A, and Allard N F 1995, {\sl AJ} {\bf 449} 258
\item Gianninas A, Bergeron P, and Fontaine G 2006 {\sl AJ} {\bf 132} 831
\item Holberg J B, and Bergeron P 2006 {\sl AJ} {\bf 132} 1221
\item Landolt A U 1992 {\sl AJ} {\bf 104} 340
\item Leggett S K et al. 2006 {\sl MNRAS} {\bf 373} 781
\item Lemke M 1997 {\sl A\&A Suppl} {\bf 122} 285 
\item Roche P F et al. 2003 {\sl SPIE} {\bf 4841} 901
\item Tokunaga A T, Simons D A, and Vacca W D 2002 {\sl PASP} {\bf 114} 180

\end{thereferences}

\end{document}